\documentclass[a4paper,12pt]{article}

\usepackage{graphicx,amsmath,amsfonts}

\begin{document}

\vspace*{0.5cm}

\begin{center}

{\Large \bf
ON THE SPECIFIC FEATURES OF TEMPERATURE EVOLUTION \\
IN ULTRACOLD PLASMAS

}

\bigskip
\bigskip

{\large \bf
Yu. V. Dumin}

\bigskip

{\it
N.V.~Pushkov Institute of Terrestrial Magnetism, Ionosphere, \\
and Radio Wave Propagation, Russian Academy of Sciences \\
IZMIRAN, Troitsk, Moscow region, 142190 Russia

\bigskip

E-mail: {\tt dumin@yahoo.com, dumin@izmiran.ru}
}

\bigskip
\bigskip

\end{center}

A theoretical interpretation of the recent experimental studies
of temperature evolution in the course of time in the freely-expanding
ultracold plasma bunches, released from a magneto-optical trap,
is discussed. The most interesting result is finding the asymptotics
of the form $ T_e \propto t^{-(1.2 \pm 0.1)} $ instead of $ t^{-2} $,
which was expected for the rarefied monatomic gas during inertial
expansion. As follows from our consideration, the substantially
decelerated decay of the temperature can be well explained by
the specific features of the equation of state for the ultracold
plasmas with strong Coulomb's coupling, whereas a heat release
due to inelastic processes (in particular, three-body recombination)
does not play an appreciable role in the first approximation.
This conclusion is confirmed both by approximate analytical estimates,
based on the model of ``virialization'' of the charged-particle
energies, and by the results of {\it ab initio} numerical simulation.
Moreover, the simulation shows that the above-mentioned law of
temperature evolution is approached very quickly---when the virial
criterion is satisfied only within a factor on the order of unity.

\bigskip

% 52.25.Kn - Thermodynamics of plasmas
% 52.27.Gr - Strongly-coupled plasmas
% 52.65.Yy - Molecular dynamics methods

\newpage

\section{Introduction}
\label{Intro}

Apart from the previously-known types of nonideal plasmas~[1],
an active study of one more kind of the nonideal Coulomb's
systems---the bunches of very rarefied ultracold plasmas
created by laser cooling and ionization in the magneto-optical
traps---was started in the recent years ({\it e.g.,}
review~[2]). They are the classical (non-quantum) gaseous systems
with a characteristic temperature of a fraction to a few Kelvin,
whose Coulomb's coupling parameter $ \Gamma = e^2 n^{1/3} / k_B T $
can reach considerable values. For example, the immediately measured
values for ions~$ {\Gamma}_i $ are~$ 2{\div}3 $~[3];
the corresponding estimates of~$ {\Gamma}_e $ for electrons are
less accurate and model-dependent, but they also give the values
comparable to unity.

The possibility of existence of such metastable plasma states
was theoretically predicted many years ago (see, for example,
article~[4] and references therein), but they were created
experimentally only after a sufficient development of the laser
cooling technique~[5, 6]. In the most recent time, similar
systems began to be studied also by gas-dynamic cryogenic
installations~[7]. Besides, creation of the same plasma states
by artificial release of gaseous clouds from spacecraft was
discussed long time ago~[8, 9]. Unfortunately, the diagnostic
possibilities in space still remain too limited to give reliable
conclusions about the properties of the resulting plasmas.

One of the most interesting results of the laboratory experiments
performed by now was studying a temporal behavior of the temperature
in the ultracold plasma clouds released from a magneto-optical trap
and expanding freely in space. It was found, firstly, that the law
of decay of the electron temperature at large times becomes
``universal'', {\it i.e.} independent of the initial conditions.
(For example, even when the initial temperatures were scattered by
30~times, the values of~$ T_e $ after a few microseconds deviate
from each other by only $ 2{\div}3 $ times, and even less later~[10].)
Secondly, which is more interesting, the measured asymptotics had
the form $ \, T_e \propto t^{-(1.2 \pm 0.1)} \approx t^{-1} $~[11]
instead of $ t^{-2} $, which should be expected for the ideal
rarefied gas without the internal degrees of freedom
($ \gamma \! = \! 5/3 $) at the inertial stage of expansion
({\it i.e.} when the plasma cloud expands with a constant rate,
so that its size increases linearly in time, $ R(t) \propto t $).

The most evident way to explain the substantially decelerated decay
in the electron temperature is to take into account a heat release
due to recombination of the charged particles. In the particular case
of atomic ions, the most efficient channel is the three-body process,
$ A^+ + e + e \to A + e $, when one electron is captured by the ion,
and the second electron carries away the excessive energy.
Unfortunately, the recent attempts of quantitative modeling
the observed law of temperature variation due to the heat release
by the tree-body recombination were unsuccessful:
the resulting dependence~$ T_e(t) $ differed only slightly from
the adiabatic case~$ t^{-2} $ (see, for example, the inset to Fig.3a
in paper~[11]%
\footnote{
Let us mention that the method of drawing the plots in paper~[11]
is somewhat confusing. According to the physical sense of the problem,
the various laws of evolution should be confronted with each other
starting from the same initial temperature . Unfortunately, the plots
in Fig.~3a of the above-mentioned paper are taken at the same
``final'' temperature (defined in some arbitrary instant of time).
Most probably, this was done just to avoid merging the curves
with a horizontal axis. As a result, it looks at the first glance
that the theory considerably disagrees with the experiment at small
rather than large times. On the other hand, it is written in the
text of the article about the disagreement at large times,
as should be expected from the physical formulation of the problem.
}).

The aim of the present article is to show that all the experimentally
measurable features in the temperature evolution (namely, both
the appearance of asymptotics almost independent of the initial
conditions and its particular form, close to~$ t^{-1} $) can be
naturally explained by the model of ``virialization'' of
the charged-particle energies, {\it i.e.} actually by changing
the equation of state of ultracold plasma for the case of strong
interparticle Coulomb's interaction. As a result, it becomes
unnecessary in the first approximation to take into account
any inelastic processes, such as the heat release by the three-body
recombination.%
\footnote{
Some theoretical arguments regarding suppression of the recombination
in ultracold plasmas can be found, for example, in paper~[12] and
references therein. Anyway, even if the recombination is taken into
account by the standard way~[11], its influence on the general law
of temperature evolution is quite small.
}

\section{Analytical Estimates}
\label{AnalEst}

We shall present below the estimates of the law of temperature
evolution during expansion of cold nonideal plasmas that were
actually performed over 10~years ago, before the experimental
measurement of temperature in the magneto-optical traps~[8, 9].
These estimates were done for some kinds of the plasma outbursts
important in astrophysical applications. We are not going
to discuss here these applications but would like to remind
the basic calculations and the results obtained.

First of all, since a kinetic energy of thermal motion of the monatomic
ideal gas during its inertial expansion decreases as~$ t^{-2} $, while
the absolute value of potential energy as~$ t^{-1} $, it is reasonable
to expect that these quantities will become equal to each other at
some instant of time, and next the plasma will evolve in the strongly
nonideal regime. In such a case, let us consider a sufficiently small
(but macroscopic) plasma volume where thermodynamic equilibrium is
believed to be established and which, therefore, can be described by
the multiparticle distribution function of the following general form:
\begin{align}
& f( \textbf{r}_1, \dots , \textbf{r}_{N_e},
\textbf{v}_1, \dots , \textbf{v}_{N_e} ) =
A_f \exp \! \Big\{
  \!\! - \! \frac{1}{k_B T_e} \, \times
\nonumber \\
& \quad \times \Big[
  \, \sum_{n=1}^{N_e} \, \frac{ m_e \textbf{v}_n^2 }{ 2 }
  + \, U( \textbf{r}_1, \dots , \textbf{r}_{N_e},
  \textbf{R}_1, \dots , \textbf{R}_{N_i} )
\Big] \! \Big\} \, ,
\label{distr_fun}
\end{align}
where $ {\bf r}_n $ and $ {\bf v}_n $ are the electron coordinates
and velocities, $ {\bf R}_n $ are the ion coordinates, and $ A_f $ is
the normalization factor. Since the kinetic energy of ions in
the laboratory experiments is much less than the electron kinetic
energy, it will be ignored here; and, therefore, the electron motion
will be treated at a fixed ion distribution. (If necessary,
the same consideration can be easily performed when all kinds of
the particles are taken into account.)

Although the potential energy~$U$ in the regime of strong Coulomb's
interaction has a very complex form and cannot be treated anymore
as a small correction to the kinetic energy, the calculation of
average value of some macroscopic quantity~$F$ depending only on
the velocities~$ {\bf v}_n $ ({\it e.g.} kinetic energy) can be
performed quite easily:
\begin{equation}
\langle F(\textbf{v}) \rangle = \frac{\displaystyle %
\int \! F(\textbf{v}) \exp \Big\{ \!\! - \! \frac{1}{k_B T_e}
  \Big[ \, \sum_{n=1}^{N_e} \, \frac{ m_e \textbf{v}_n^2 }{ 2 }
  \Big] \Big\} \, d \textbf{v} }{\displaystyle %
\int \! \exp \Big\{ \!\! - \! \frac{1}{k_B T_e} \Big[ \,
  \sum_{n=1}^{N_e} \, \frac{ m_e \textbf{v}_n^2 }{ 2 }
  \Big] \Big\} \, d \textbf{v} } \; ,
\label{gen_avr}
\end{equation}
because the integrals~$ \int \exp \{ - U (\textbf{r} ,
\textbf{R}) / k_B T_e \} \, d \textbf{r} $ in the numerator and
denominator automatically cancel each other. (For conciseness,
$ \textbf{v} $, $ \textbf{r} $, and $ \textbf{R} $ denote here
the sets of all velocities and coordinates of the electrons and
ions, respectively.)

Particularly, when the kinetic energy of an ensemble of particles
is calculated, the integral in the numerator of formula~(\ref{gen_avr})
is reduced to the combination of Gaussian exponents, whose method of
calculation is well known. By such a way, it can be shown that
the average kinetic energy per one particle is given by exactly
the same expression as for ideal gas:
\begin{equation}
\langle k \rangle = (3/2) \, k_B T_e \: ;
\label{kin_en}
\end{equation}
although, let us mention once again, it is valid for the plasma with
arbitrarily strong Coulomb's interaction.

Unfortunately, if we need to calculate the average values of
quantities which are the functions of coordinates ({\it e.g.}
the average potential energy), then the distribution
function~(\ref{distr_fun}) becomes actually useless, because
the integrals involving the potential energy~$ U $ in the exponent
cannot be calculated in any reasonable approximation. Nevertheless,
a special ``way around'' can be used here. Namely, let us relate
the average potential energy per one particle~$ \langle u \rangle $
to the average kinetic one~$ \langle k \rangle $ by the virial
theorem for the Coulomb's field~[13], which is also valid at any
strength of the interparticle interaction:%
\footnote{
Let us mention that the virial theorem formulated for macroscopic
bodies in some textbooks on statistical physics (see, for example,~[14])
involves an additional term of the form $ 3 P V $, where $P$~is
the pressure, and $V$~is the volume of the system. This term appears
due to the surface integral for the particles interacting with a wall,
where the potential energy is no longer the Euler homogeneous function.
Since, in the case under consideration, the plasma is not confined
by the walls, and the potential energy of interaction between its
particles is everywhere the Euler homogeneous function, then
no extra terms should appear in the virial theorem.}
\begin{equation}
\langle k \rangle = (1/2) \: | \langle u \rangle | \: .
\end{equation}
Of course, it is necessary to assume here the ergodicity
of the system, {\it i.e.} that the quantities averaged over
a statistical ensemble are equal to the ones averaged over time.

Let us mention also that a necessary condition to apply
the virial theorem is a bounded phase volume of the system,
{\it i.e.} the particles must move in a limited spatial region
with the velocities limited by the absolute value.
The first condition, strictly speaking, is not satisfied for
the plasma cloud infinitely expanding in space. Nevertheless,
one can expect that the virial theorem can be reasonably
applicable to the system whose phase volume is unbounded
but increases with a small rate as compared to the characteristic
velocities of its particles. This condition is well satisfied
in the experiments with ultracold plasmas, because
the characteristic time of variation in macroscopic parameters
of the cloud ($ \sim \! 10^{-5} $~s for the particular
experimental setup~[11, 15]) is much greater than the periods
of microscopic motion of the electrons ($ 10^{-9} \div 10^{-7}$~s).
We shall discuss the problem of applicability of the virial
approximation in more detail in the end of the article.

At the last step of our estimates, the average potential
energy can be evidently expressed through the characteristic
distance between the particles or the plasma density:
\begin{equation}
\langle u \rangle \sim \, e^2 \! / \langle r \rangle \sim \,
e^2 n^{1/3} \, .
\label{cul_en}
\end{equation}

Finally, combining the formulas~(\ref{kin_en})--(\ref{cul_en}),
we get $ T_e \! \propto n^{1/3} $. In particular, if the cloud
expansion is inertial ({\it i.e.} linear in time) and, consequently,
its concentration changes as~$ t^{-3} $, then
\begin{equation}
T_e \! \propto t^{-1} \, .
\end{equation}

Therefore, the presented model of ``virialization'' of the charged-particle
energies in the regime of strong Coulomb's coupling well explains
the both experimental features, namely:\\
(a)~the system ``forgets'' in the course of time about its initial
temperature, {\it i.e.} the plasma clouds with various initial temperatures
begin to evolve similarly; and\\
(b)~the particular form of the time dependence is close to~$ t^{-1} $
instead of~$ \, t^{-2} $, expected intuitively. The heat release due to
inelastic processes (particularly, three-body recombination) is not
of importance here.

\section{Numerical Simulation}
\label{CompMod}

\subsection{Formulation of the Model}
\label{FormMod}

To verify the above analytical estimates, we performed {\it ab initio}
numerical simulation, based on the solution of the equations of
classical mechanics for the multiparticle system. Our approach
differs from the preceding works by the following items.

The authors of the most of simulations of ultracold plasmas performed
by now tried to include in their calculations as many particles
as possible. Consequently, to keep the computational time in reasonable
limits, they used a substantial simplification of the Coulomb's
interactions, namely:\\
(1)~to avoid large integration errors during the close collisions,
the Coulomb's potential was cut off of smoothed out at the small
distances ({\it e.g.} $ 1/r $ was changed to $ 1 / (r + r_0) $ or
something like that)~[12, 16, 17];\\
(2)~to improve convergence of the Coulomb sums at large distances,
the researchers used the multipole-tree method~[18] and various
approximations for the equations of motion of the light particles
(electrons), such as the particles-in-cell (PIC) method~[19] or
``Vlasov approximation'' for the electron component of plasma~[20]
({\it i.e.,} actually, the introduction of self-consistent field,
ignoring the interparticle correlations).

All these approximations distort the functional dependence of
the Coulomb's potential so strongly that one can hardly expect
the virial relations to be satisfied, because the virial theorem
is very sensitive to the particular form of the potential energy.

As distinct from the above approaches, we tried in our simulation
to integrate the equations of motion of the charged particles in
the real electric microfields as accurately as possible, without any
artificial distortion. With this aim in view, we used the basic cell
with a relatively small number of particles ({\it e.g.,} a few
dozens) which was assumed to be supplemented in all directions
by infinite number of mirror cells. Coulomb forces acting on
each particle in the basic cell were calculated taking into account
not only the particles in the same cell but also in all its
mirror images, until the specified accuracy is achieved (Fig.~1).

%%%%%%%%%%%%%%%%%%%%%%%%%%%%%%%%%%%%%%%%%%%%%%%%%%%%%%%%%%%%%%%%%%
\begin{figure}[t]
\centerline{
\includegraphics[width=9cm]{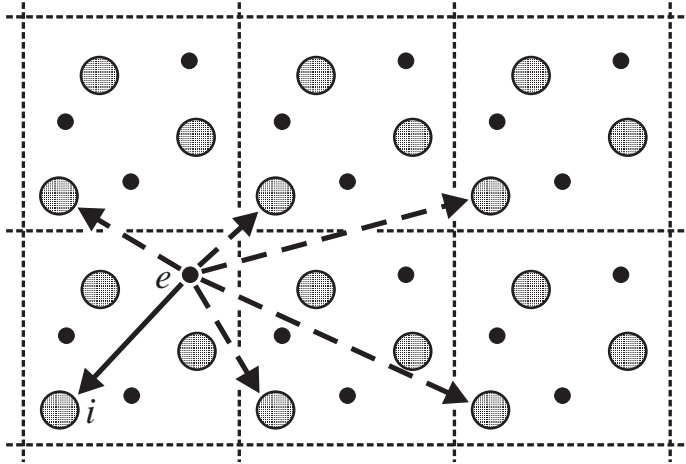}
}
\caption{\sl
Sketch illustrating the interaction between an electron ($ e $)
and ion ($ i $) both inside the basic cell (solid arrow) and with all
its mirror images (dashed arrows). The ions are pictorially drawn by
the large circles; and electrons, by the small ones (in fact,
all the particles in our simulation were point-like).}
\end{figure}
%%%%%%%%%%%%%%%%%%%%%%%%%%%%%%%%%%%%%%%%%%%%%%%%%%%%%%%%%%%%%%%%%%

So, we worked with a relatively small number of the equations of
motion---only for particles in the basic cell. As a result,
we could perform integration with a very small step and, therefore,
automatically avoid the problem of close collisions without any
artificial modification of the Coulomb's potential. On the other hand,
the total number of the charged particles participating in Coulomb
interactions (including the mirror images) was in our calculations
between 100\,000 and 3\,000\,000. This is not less and even usually
much greater than in the simulations by other researchers.

\subsection{Initial Equations}
\label{InitEq}

As was already mentioned in Sec.~\ref{AnalEst}, the ion kinetic energy
in the experiments with magneto-optical traps is usually small as compared
to the kinetic energy of electrons. Consequently, the ions in our
simulation were assumed to move from the very beginning by the inertial
({\it i.e.,} linear in time) law; while the electron dynamics was
described taking into account Coulomb's interactions not only inside
the basic cell but also with infinite number of its mirror images:
\begin{eqnarray}
&& m \, \frac{d^2}{dt^2} \, {\bf r}_i = \,
  \sum_{j=1}^{N} Z e^2 \,
    \frac{{\bf R}_j - {\bf r}_i}{|{\bf R}_j - {\bf r}_i|^3} \, + \,
  \sum_{j=1, \, j \neq i}^{ZN} e^2 \,
    \frac{{\bf r}_i - {\bf r}_j}{|{\bf r}_i - {\bf r}_j|^3} +
\nonumber \\
&& + \,
  \sum_{k=1}^{3} \: \sum_{n_k = -\infty, \, n_k \neq 0}^{+\infty}
    \Bigg\{
      \sum_{j=1}^{N} Z e^2
        \frac{[\, {\bf R}_j + L \sum_{l=1}^{3} n_l {\bf e}_l \,] - {\bf r}_i}%
        {|[\, {\bf R}_j + L \sum_{l=1}^{3} n_l {\bf e}_l \,] - {\bf r}_i|^3} +
\nonumber \\
&& + \,
      \sum_{j=1}^{ZN} \, e^2
        \frac{{\bf r}_i - [\, {\bf r}_j + L \sum_{l=1}^{3} n_l {\bf e}_l \,]}%
        {|{\bf r}_i - [\, {\bf r}_j + L \sum_{l=1}^{3} n_l {\bf e}_l \,]|^3}
    \Bigg\} \: .
\label{eq:electron_motion}
\end{eqnarray}
Here, ${\bf R}_i \; (i = 1,..., N)$~are the ion coordinates,
${\bf r}_i \; (i = 1,..., ZN)$~are the electron coordinates,
$N$~is the number of ions in the basic cell, $Z$~is the ion charge
(in the calculations presented below in this article, the ions were
singly charged, {\it i.e.} $ Z \equiv 1 $), $m$ and $e$~are the mass
and absolute value of the electron charge, $L$~is the linear size of
the basic cell, ${\bf e}_l \; (l = 1, 2, 3)$~are the unit vectors of
the Cartesian coordinate system, and subscripts~$n_k \; (k = 1, 2, 3) $
specify the number of the mirror cell in each direction.
Let us mention that a software code actually performed summation over
the mirror cells not from~$ - \infty $ to ~$ + \infty $, as in
formula~(\ref{eq:electron_motion}), but starting from the central
(basic) cell, one shell of cells after another, until a specified
criterion of convergence of the Coulomb sums is satisfied.

Yet another well-known problem in modeling the freely-expanding
plasmas is a considerable variation in the spatial scale of the system
(and, consequently, in the values of Coulomb's forces) during its
evolution. As a result, it becomes difficult to choose the method of
numerical integration ensuring a stable accuracy for the entire solution.
In our modeling, this problem was resolved by introduction of
a ``scalable'' coordinate system, expanding in space with a mean plasma
expansion velocity. In other words, size of the basic cell was taken
to be increasing linearly in time:
\begin{equation}
L = L_0 + u_0 \, t
\label{eq:norm_length_with_time}
\end{equation}
(which corresponds, from the physical point of view, just to the inertial
motion), and the coordinates of all particles were normalized to this
time-dependent scale.

The ions, moving by inertia, will be exactly at rest in such coordinates,
while the electrons will move within a cell of the fixed size.
When the electron crosses one of boundaries of the cell, it is assumed
to appear at the opposite side, as it is usually done in the method of
molecular dynamics.

At last, to get the final equations of motion for the electrons,
which will be used in the simulation, all physical quantities should
be reduced to the dimensionless form. Let the time-dependent unit of
length~$ \tilde{l} \, $ be the characteristic interparticle distance,
determined by the following way. The total number of particles in
the basic cell equals $ (Z+1)N $; so the volume per one particle
will be $ {L^3}/((Z+1)N) $. Consequently, the characteristic linear
size can be determined as
\begin{equation}
\tilde{l} = \frac{L}{(Z+1)^{1/3} \: N^{1/3}} \: .
\label{eq:norm_length}
\end{equation}
Particularly, at the initial instant of time (which from here on
will be denoted by subscript~0) we get:
\begin{equation}
\tilde{l_0} = \frac{L_0}{(Z+1)^{1/3} \: N^{1/3}} \: .
\label{eq:norm_length_0}
\end{equation}

The characteristic time scale~$ \tau $ can be formally introduced,
for example, by the virial relation taken at the initial instant
of time:
$ (1/2) \, m \, (\tilde{l_0} / \tau)^{\, 2} \, = \,
(1/2) \, Z e^2 / \, \tilde{l_0} \, $,
from which we obtain
\begin{equation}
\tau = {\left( \frac{m}{Z e^2} \right)}^{\! 1/2} \:
       {\tilde{l_0}}^{\, 3/2} .
\label{eq:norm_time}
\end{equation}
Within a numerical factor, this coincides with the inverse
Langmuir frequency (as could be expected from the dimensionality
arguments).

From here on, the quantities normalized to~$ \tilde{l} $ and
$ \tau $ will be denoted by asterisks. It can be easily shown that
the physical scale~$ \, \tilde{l} \, $ will be a function of
the dimensionless time of the form:
\begin{equation}
\tilde{l} = \, \tilde{l_0} \, ( 1 + \, u_0^* \, t^* ) \, ,
\label{eq:L_dimles_time_dimles}
\end{equation}
where
\begin{equation}
u_0^* = u_0 \, \tau / L_0 \, ;
\label{eq:u_0_dimles_def}
\end{equation}
while a dimensionless length of the basic cell is constant:
\begin{equation}
L^* = ( Z + 1 )^{1/3} N^{1/3} .
\label{eq:dimles_L}
\end{equation}

Therefore, in the variables specified above, equations of the
electron motion~(\ref{eq:electron_motion}) will take the form:
\begin{equation}
{\ddot{\rm \bf r}}_i^*
+ \, 2 \, u_0^* \, ( 1 + \, u^*_0 \, t^* )^{-1} \,
  {\dot{\rm \bf r}}_i^* = \,
( 1 + \, u^*_0 \, t^* )^{-3} \, {\rm \bf F}^*_i \, ,
\label{eq:el_motion_compact}
\end{equation}
where dot denotes a derivative with respect to the dimensionless
time~$ t^* $, and~$ {\rm \bf F}^*_i $~is the total Coulomb's force
acting on~$i$th electron from all other electrons and ions:
\begin{eqnarray}
&&
{\rm \bf F}^*_i = \,
  \sum_{j=1}^{N}
    \frac{{\bf R}^*_j - {\bf r}^*_i}{|{\bf R}^*_j - {\bf r}^*_i|^3} \, + \,
\frac{1}{Z} \!
  \sum_{j=1, \, j \neq i}^{ZN}
    \frac{{\bf r}^*_i - {\bf r}^*_j}{|{\bf r}^*_i - {\bf r}^*_j|^3} \, +
\nonumber \\[2mm]
&&
+ \, \sum_{k=1}^{3} \: \sum_{n_k = -\infty, \, n_k \neq 0}^{+\infty}
    \Bigg\{
      \sum_{j=1}^{N}
      \frac{[\, {\bf R}^*_j + L^{{}^{\scriptstyle *}}
        \sum_{l=1}^{3} n_l {\bf e}_l \,] - {\bf r}^*_i}%
      {|[\, {\bf R}^*_j + L^{{}^{\scriptstyle *}}
        \sum_{l=1}^{3} n_l {\bf e}_l \,] - {\bf r}^*_i|^3} \, +
\nonumber \\[2mm]
&&
+ \, \frac{1}{Z} \, \sum_{j=1}^{ZN}
      \frac{{\bf r}^*_i - [\, {\bf r}^*_j + L^{{}^{\scriptstyle *}}
        \sum_{l=1}^{3} n_l {\bf e}_l \,]}%
      {|{\bf r}^*_i - [\, {\bf r}^*_j + L^{{}^{\scriptstyle *}}
        \sum_{l=1}^{3} n_l {\bf e}_l \,]|^3} \Bigg\} \, .
\label{eq:dimles_force}
\end{eqnarray}

As follows from equation~(\ref{eq:el_motion_compact}), the effect of
inertial plasma expansion in the ``expanding'' coordinate system
looks like the influence of an effective dissipative force, which is
proportional to the electron velocities. Therefore, temperature
of the electron gas results from the balance of two effects---on
the one hand, acceleration and heating of the electrons due to
Coulomb's interactions and, on the other hand, their deceleration and
cooling by the above-mentioned dissipative forces.

\subsection{Method of Computation}
\label{SolMeth}

To write conveniently the subsequent formulas, let us introduce
the auxiliary quantity:
\begin{equation}
s(t^*) = \, ( 1 + \, u^*_0 \, t^* )^{-1} \, .
\label{eq:scale_factor}
\end{equation}

Then, following the standard procedure, the second-order
equation~(\ref{eq:el_motion_compact}) can be rewritten as a set of
two equations of the first order:
\begin{subequations}
\begin{eqnarray}
&& {\dot{\rm \bf r}}_i^* = \, {\rm \bf v}_i^* \, ,
\label{eq:diff_eq1}
\\
&& {\dot{\rm \bf v}}_i^* =
   - 2 \, u_0^* s \, {\rm \bf v}_i^* + \,
   s^3 \, {\rm \bf F}^*_i \, ,
\label{eq:diff_eq2}
\end{eqnarray}
\end{subequations}
which can be solved by any available method of numerical integration.
We used Runge--Kutta method of the second order.

The set of equations~(\ref{eq:diff_eq1}) and~(\ref{eq:diff_eq2})
is solved in the region
\begin{equation}
- L^* \! / 2 \: \leq {(r^*_i)}_l \: \leq L^* \! / 2 \, ,
\quad \mbox{where \:} i = 1,..., ZN \;\, \mbox{and \;} l = 1,2,3
\label{eq:r_electr_interval}
\end{equation}
(the subscript~$l$ denotes the number of the Cartesian coordinate)
with standard molecular-dynamic boundary conditions:
a particle leaving the basic cell of simulation through one of
its sides is replaced by the particle entering the cell with
the same velocity through the opposite side. The initial coordinates
of electrons are given by the random-number generator as a uniform
statistical distribution over the region~(\ref{eq:r_electr_interval});
and the initial velocities, as Gaussian distribution with a specified
dispersion.

As was already mentioned above, an ion motion in the simulated
experimental conditions can be considered as inertial; so that
the normalized ion coordinates in the ``expanding'' reference system
remain constant. At the initial instant of time, they are taken as
a uniform statistical distribution over the region
\begin{equation}
- L^* \! / 2 \: \leq {(R^*_i)}_l \: \leq L^* \! / 2 \, ,
\quad \mbox{where \:} i = 1,..., N \;\, \mbox{and \;} l = 1,2,3 \; .
\label{eq:r_ion_interval}
\end{equation}
It is unnecessary, of course, to specify the initial velocities
and to solve the equations of motion for the ions.

Finally, let us discuss in more detail how to calculate a kinetic energy
of the electron motion~$ K $ with respect to the mean plasma flow,
which is related to their temperature by formula~(\ref{kin_en}). We can
see here one more advantage of the introduced coordinate system:
the required kinetic energy is calculated in this system just by
differentiating the normalized (dimensionless) electron coordinates with
respect to time; and one should not differentiate the normalization
factor~$ \, \tilde{l}(t) \, $ itself, because its variation is associated
with a kinetic energy of the plasma motion as a whole:
\begin{equation}
K = \, \frac{m}{2} \, \sum_{i=1}^{ZN}
  {\bigg( \frac{d}{dt} \, {\rm \bf r}_i \! \bigg)}_{\rm \! rel}^{\! 2} = \,
\frac{m \, {\tilde{l}}^{\, 2}}{2 \, {\tau}^2} \, \sum_{i=1}^{ZN}
  {\bigg( \frac{d}{dt^*} \, {\rm \bf r}^*_i \! \bigg)}^{\!\! 2} = \,
\frac{m \, {\tilde{l}}^{\, 2}}{2 \, {\tau}^2} \, \sum_{i=1}^{ZN} \,
( {\rm \bf v}^*_i \! )^{\, 2} \, .
\label{eq:Kin_ener_def}
\end{equation}

By introducing the normalization factor
\begin{equation}
\tilde{K} = \, \frac{m}{2} \, \frac{{\tilde{l_0}}^2}{{\tau}^2} \; ,
\label{eq:Kin_ener_norm_fac}
\end{equation}
we find that the dimensionless kinetic energy of the relative motion
is given by the expression:
\begin{equation}
K^* =
\, s^{-2} \sum_{i=1}^{ZN} \, ( {\rm \bf v}^*_i \! )^{\, 2} \, .
\label{eq:Kin_ener_dimles}
\end{equation}
Just this formula (after division by the total number of particles
in the basic cell) will be used below to calculate the plasma temperature.

\subsection{Particular Case of the Ideal Gas}
\label{IdealGas}

Before presentation of the results of numerical modeling, let us discuss
one particular example, which can be completely solved in analytic form
and demonstrates a self-consistency of the approach used. Namely, let us
consider the case of an ideal gas, when the forces of interparticle
interaction~$ {\rm \bf F}^*_i $ disappear at all (for example, because
the electric charges tend to zero). Then, equation~(\ref{eq:diff_eq2})
is reduced to
\begin{equation}
{\dot{v}}_{ik}^* = - 2 \, u_0^* s \, v_{ik}^*
\label{eq:eq_motion_ideal}
\end{equation}
(subscript~$i$ denotes here the number of particle, and $k$~is the number
of coordinate). After the integration of this differential equation,
taking into account the particular form of function~$ s(t^*) $ given by
formula~(\ref{eq:scale_factor}), we get:
\begin{equation}
v_{ik}^* = (v_{ik}^*)_{{}_{\scriptstyle 0}} \, ( 1 + \, u^*_0 \, t^* )^{-2} =
(v_{ik}^*)_{{}_{\scriptstyle 0}} \, s^2(t^*) \, .
\label{eq:Solution_ideal}
\end{equation}

At last, substituting solution~(\ref{eq:Solution_ideal}) to general
expression for the kinetic energy~(\ref{eq:Kin_ener_dimles}), we find:
\begin{equation}
K^* =
\, s^{-2} \sum_{i=1}^{ZN} \, ( {\rm \bf v}^*_i \! )_{0}^{\, 2} \, s^4 =
\, K^*_0 \: s^2 (t^*) \: .
\label{eq:Kin_ener_ideal_gas}
\end{equation}
Therefore, the electron temperature $ \, T_e \propto K \propto K^* \, $
will change in time by the law:
\begin{equation}
T_e(t) \propto s^2(t^*) \propto ( 1 + \, u^*_0 \, t^* )^{-2}
       \propto t^{-2}
\quad \mbox{at large } t \, ,
\label{eq:Temp_time}
\end{equation}
as should be expected for the inertial expansion of the ideal monatomic gas.

\subsection{Basic Parameters of the Numerical Simulation}
\label{ModPar}

Before presentation of the results of numerical simulation for
strongly-coupled plasmas, let us describe in more detail the basic
parameters used in the computation presented below in Figs.~2 and 3.

{\bf The number of particles in the basic cell:}
$ N_{\rm tot} = (Z+1) \, N = 20 \, $ ({\it i.e.,} 10~electrons and
10~ions). Due to the so small number of particles, we were able to perform
integration with high accuracy (see below for more details) over the time
interval $ {\Delta}_{\rm tot} t^* \approx 1000 $. The total number of
particles taken into account in the calculation of Coulomb sums
(including the mirror cells) was, on the average, a few hundred thousand
(more exactly, 137\,180 to 3\,327\,500).

In other versions of simulation, for example, when the number of particles
in the basic cell was increased by an order of magnitude---up
to~$ \, N_{\rm tot} = 200 $, we were enforced to decrease the interval
of integration by $ 3{\div}4 $~times; but the behavior of macroscopic
plasma parameters remained qualitatively the same. When $ N_{\rm tot} $ was
further increased by a few more times, the reachable interval of
integration became so short that it was difficult to draw reliable
conclusions on the laws of evolution of the plasma parameters with time.

{\bf The step of integration and the accuracy of calculation of
the Coulomb's forces:}
$ \Delta t^* = 10^{-3}{\div}\,10^{-4} $ and
$ {\varepsilon}_F = 10^{-3}{\div}\,10^{-4} $.
These values were chosen empirically, by a series of test simulations.
Starting from the above-written values, the behavior of macroscopic
plasma parameters became reproducible, {\it i.e.} independent of
further increase in the accuracy of computation. Let us mention that
the integration step $ \Delta t^* = 10^{-3} $ was usually sufficient.
However, at some ``unfavorable'' initial distributions of the particles,
there might be a few close collisions when such integration step was
too large.%
\footnote{
The insufficiently small value of the integration step in close collisions
manifests itself, first of all, as a ``sling effect'', {\it i.e.} a sudden
ejection of the electron during a passage of pericenter of the orbit.
In macroscopic plasma description, such situations look as non-physical
saw-toothed peaks in the dependence of kinetic energy on time.}
In these cases, it should be reduced to~$ 10^{-4} $ and sometimes even
smaller.

{\bf Velocity of the inertial plasma expansion:} $ u_0^* = 0.1 \, $.
Such value of the expansion velocity at the time interval under
simulation results in increasing the plasma cloud size by a few
dozen times, which corresponds to the experimental conditions~[11].

{\bf Root-mean-square scatter of the initial velocities of
the particles:} $ {\sigma}_{v0}^* = 3.0 $ (in each Cartesian
coordinate), {\it i.e.} initially the plasma was taken to be
slightly nonideal. In some other simulations, we used greater
values of~$ {\sigma}_{v0}^* $, {\it i.e.} started from a more
ideal plasma state. In such cases, the initial stage of evolution
of the electron temperature was close to the expected dependence
$ T_e \propto t^{-2} $. In subsequent modeling, to avoid spending
a lot of time for the integration over the physically trivial interval,
we preferred to start just from $ {\sigma}_{v0}^* = 3.0 \, $.

\subsection{Results of the Simulation}
\label{CompRes}

The results of our numerical simulation for the electron temperature
are presented by crosses in logarithmic scale in Fig.~2. After excluding
from a consideration the earliest period of time, when relaxation
processes occurred, the points in the remaining time interval lie almost
along a straight line, corresponding to the power law
$ T_e \propto t^{\alpha} $. The respective exponent was found to be
in the range $ \alpha = - (1.08{\div}1.25) $. This is quite close to
the value $ \alpha = - 1 $, following from the simple virial estimate,
and is in perfect agreement with the experimental value
$ \alpha = - (1.1{\div}1.3) $~[11].

%%%%%%%%%%%%%%%%%%%%%%%%%%%%%%%%%%%%%%%%%%%%%%%%%%%%%%%%%%%%%%%%%%
\begin{figure}[t]
\centerline{
\includegraphics[width=10cm]{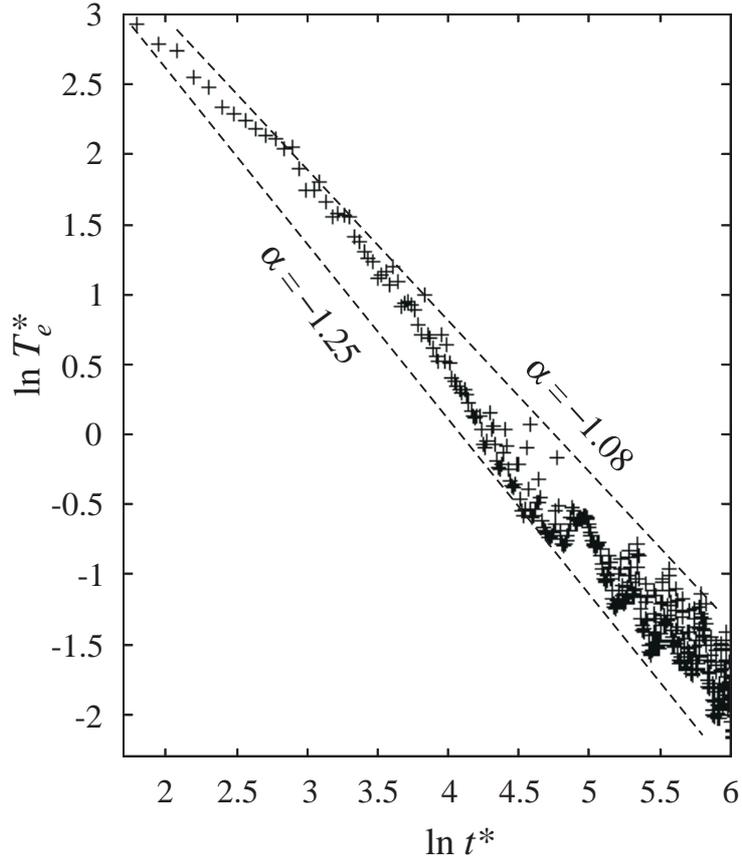}
}
\caption{\sl
Electron temperature obtained by the numerical simulation
as function of time in the logarithmic coordinate system.
The inclined straight lines show the power-like dependences
of the form~$ \, T_e \propto t^{\alpha} $; the values of~$ \alpha $
being presented near the respective lines.}
\end{figure}
%%%%%%%%%%%%%%%%%%%%%%%%%%%%%%%%%%%%%%%%%%%%%%%%%%%%%%%%%%%%%%%%%%

It is reasonable to ask if the dependence obtained is really caused by
the effect of virialization, discussed in Sec.~\ref{AnalEst}?
To answer this question, we plotted in Fig.~3 the temporal dependences
of average kinetic energy and one-half the absolute value of
the potential (Coulomb's) energy%
\footnote{
Let us mention that the value of Coulomb's energy used here is not just
an estimate by the characteristic interparticle distance, but it is
the quantity accurately calculated by the summation over all particles,
taking into account the mirror cells, as in the case of Coulomb's forces.
}
in ordinary (not logarithmic) coordinates. These quantities were
obtained by the method of moving average over the interval
$ {\Delta}_{\rm avr} t^* = 20 $ ({\it i.e.} 10~units of dimensionless
time in both sides from the center). It is seen in this figure that
at large time, $ t^* \approx 500 $, the curves
$ \overline{k}^{\, *}(t^*) $ and $ | \overline{u}^{\, *}(t^*) | / 2 $
begin to coincide with a sufficiently good accuracy, about~{10\%}.
On the other hand, it is important to emphasize that the above-mentioned
power law of evolution of~$ T_e $ with exponent $ \alpha = - (1.1{\div}1.3) $
is established much earlier, when the virialization criterion is satisfied
only on the order of magnitude. Probably, this is just the reason
why~$ \alpha $ differs from the exact value of~$ -1 $. This question
requires a further careful study.

%%%%%%%%%%%%%%%%%%%%%%%%%%%%%%%%%%%%%%%%%%%%%%%%%%%%%%%%%%%%%%%%%%
\begin{figure}[t]
\centerline{
\includegraphics[width=11cm]{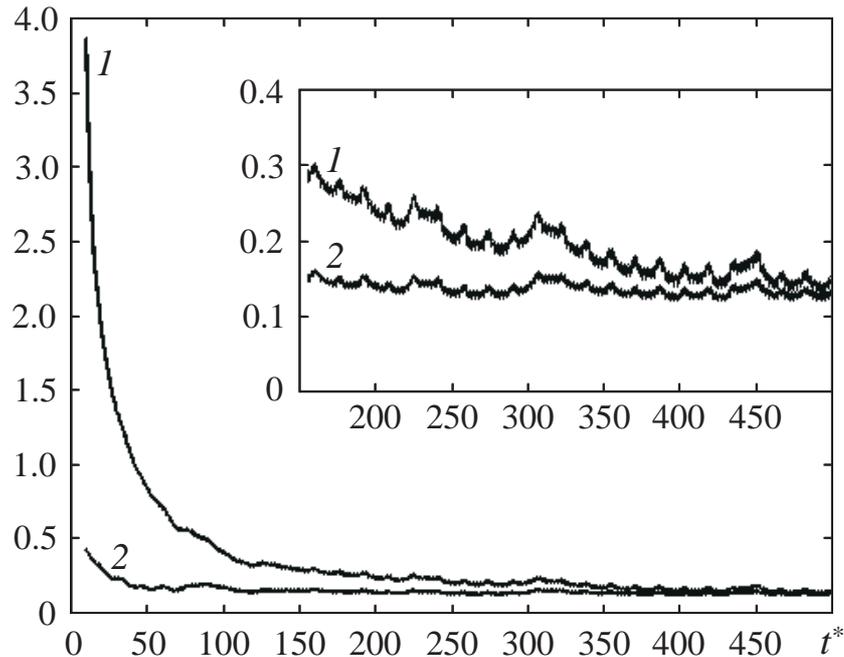}
}
\caption{\sl
Temporal behavior of the average kinetic energy per one
particle~$ \overline{k}^{\, *} $ (curve~{\it 1}) and one-half
the absolute value of average potential (Coulomb's)
energy~$ | \overline{u}^{\, *} | / 2 $ (curve~{\it 2}).}
\end{figure}
%%%%%%%%%%%%%%%%%%%%%%%%%%%%%%%%%%%%%%%%%%%%%%%%%%%%%%%%%%%%%%%%%%

At last, it is interesting to compare our findings with the results of
paper~[21], which was based on a quite complex ``hybrid model'' of
ultracold plasmas. It involved the quasi-hydrodynamic equations for
ions, description of the electron dynamics by Monte-Carlo method,
as well as additional inclusion of some collisional processes
by the {\it a priori} information about their cross-sections.
As a result, it was in particular found that, if the three-body
recombination was not included explicitly, then the electron
coupling parameter~$ {\Gamma}_e $ increases infinitely with time
at any initial temperature. On the other hand, if the effects of
three-body recombination and subsequent inelastic collisions of
electrons with the produced Rydberg atoms were included to the equations
as additional terms, then the coupling parameter was stabilized
in the course of time at the level~$ {\Gamma}_e \approx 0.2 $
(see Fig.~3 in paper~[21]).

Unfortunately, this conclusion does not agree with the results of
paper~[17], published almost at the same time. Its authors used a ``pure''
molecular-dynamic model, without any explicit inclusion of
the three-body recombination, and found a stabilization of~$ {\Gamma}_e $
at the level about unity. However, it should be mentioned that only
the slightly-expanding plasmas were considered. As regards our simulation,
it also does not take into consideration any special corrections for
the recombination and gives the asymptotic values of coupling parameter
$ {\Gamma}_e \approx 1 $, {\it i.e.} in agreement with~[17], also in
the case of very large plasma expansion (by a few dozen times, in terms of
the linear size).%
\footnote{
Comparing the values of coupling parameter~$ \Gamma $ obtained in
the various works, one should keep in mind that definitions
of this parameter may be slightly different, by a factor about
unity.}

Let us emphasize that, if the recombination processes are not
taken into account explicitly in the equations of molecular dynamics,
this does not mean that these processes are completely ignored.
In fact, our model already involves their description from
the first principles: as follows from a more careful analysis
of the computational results, the scattering of crosses in Fig.~2
and spiky-like behavior of the curves in Fig.~3 are caused just by
formation of quasi-bound states of the electrons and ions ({\it i.e.}
by the onset of recombination between the charged particles).
If the electron--ion pair with a sufficiently large eccentricity
of the orbit is formed, then every passage of the electron in
the vicinity of the ion will be associated with a sharp outburst
of both kinetic energy and the absolute value of potential energy.
This looks as a series of sharp equidistant peaks in macroscopic
parameters of the plasma. A few series of such peaks were clearly
observed in our simulations. However, consideration of this phenomenon
requires a separate article, and it will not be discussed here in
more detail.

\section{Conclusions}
\label{Concl}

\hspace{9mm}
1.~The law of evolution of the electron temperature close to~$ t^{-1} $
can be explained, in the first approximation, by a simple analytical
model based on the virialization of energy of the charged particles.

2.~The numerical simulation from the first principles, using the scalable
coordinate system and accurately taking into account Coulomb's
interactions, results in some deviation of the exponent, namely,
from~$ -1 $ to $ -(1.08{\div}1.25) $, which is in perfect agreement with
the experimental value.

3.~Therefore, the decelerated law of decrease in the electron temperature
is, first of all, a manifestation of the specific equation of state of
the cold nonideal plasma rather than a result of additional heat release
due to recombination.

4.~It was unexpectedly found in the numerical simulation that
the law of variation in the electron temperature of
the form~$ T_e \propto t^{-(1.08{\div}1.25)} $ is established very
quickly---when the virial relation for energies is satisfied only
within a factor on the order of unity.

5.~A more careful analysis of the results of our simulation reveals also
some finer effects, for example, the quasi-periodic oscillations of
energy caused by the formation of bound electron--ion states.

\section*{Acknowledgements}

The most part of numerical simulation described in the present paper
was performed by the computer cluster of Max-Planck-Institut f{\"u}r
Physik komplexer Systeme, Dresden, Germany. I am grateful to
the Head of IT Department of this institute H.~Scherrer-Paulus
for a continuous help in my work, as well as to the Director of
the institute J.-M.~Rost for the inclusion of my proposal to
the research plan. Some additional calculations were done also
in Russia by ordinary PC; and I got a considerable technical
assistance from V.A.~Koutvitsky and P.A.~Rodin.

%%%%%%%%%%%%%%%%%%%%%%%%%%%%%%%%%%%%%%%%%%%%%%%%%%%%%%%%%%%%

\bigskip
\bigskip
\bigskip
\bigskip

\noindent
{\bf \large REFERENCES:}

\medskip

1.~{\it Fortov~V.E., Khrapak~A.G., Yakubov~I.T.} \/
Physics of Nonideal Plasmas (Fizika neideal'noi plazmy),
Fizmatlit, Moscow, 2004 (in~Russian).

2.~{\it Killian~T.C., Pattard~T., Pohl~T., Rost~J.M.} //
Phys.\ Rep. 2007. V.~449. P.~77.

3.~{\it Simien~C.E., Chen~Y.C., Gupta~P., et al.} //
Phys.\ Rev.\ Lett. 2004. V.~92. P.~143001.

4.~{\it Tkachev~A.N., Yakovlenko~S.I.} //
JETP Lett. 2001. V.~73. P.~66.

5.~{\it Gould~P., Eyler~E.} //
Phys.\ World. 2001. V.~14(3). P.~19.

6.~{\it Bergeson~S., Killian~T.} //
Phys.\ World. 2003. V.~16(2). P.~37.

7.~{\it Morrison~J.P., Rennick~C.J., Keller~J.S., Grant~E.R.} //
Phys.\ Rev.\ Lett. 2008. V.~101. P.~205005.

8.~{\it Dumin~Yu.V.} //
J. Low Temp.\ Phys. 2000. V.~119. P.~377.

9.~{\it Dumin~Yu.V.} //
Astrophys.\ Space Sci. 2001. V.~277. P.~139.

10.~{\it Roberts~J.L., Fertig~C.D., Lim~M.J., Rolston~S.L.} //
Phys.\ Rev.\ Lett. 2004. V.~92. P.~253003.

11.~{\it Fletcher~R.S., Zhang~X.L., Rolston~S.L.} //
Phys.\ Rev.\ Lett. 2007. V.~99. P.~145001.

12.~{\it Lankin~A.V., Norman~G.E.} //
J. Phys.\ A: Math.\ Theor. 2009. V.~42. P.~214042.

13.~{\it Landau~L.D., Lifshitz~E.M.}
Mechanics (3rd ed.), Pergamon, Oxford, 1976.

14.~{\it Landau~L.D., Lifshitz~E.M.}
Statistical Physics, V.~1 (3rd ed.), Pergamon, Oxford, 1980.

15.~{\it Fletcher~R.S., Zhang~X.L., Rolston~S.L.} //
Phys.\ Rev.\ Lett. 2006. V.~96. P.~105003.

16.~{\it Bobrov~A.A., Bronin~S.Ya., Zelener~B.B., et al.} //
J.\ Exp.\ Theor.\ Phys. 2008. V.~107. P.~147.

17.~{\it Kuzmin~S.G., O'Neil~T.M.} //
Phys.\ Rev.\ Lett. 2002. V.~88. P.~065003.

18.~{\it Mazevet~S., Collins~L.A., Kress~J.D.} //
Phys.\ Rev.\ Lett. 2002. V.~88. P.~055001.

19.~{\it Robicheaux~F., Hanson~J.D.} //
Phys.\ Plasm. 2003. V.~10. P.~2217.

20.~{\it Pohl~T., Pattard~T., Rost~J.M.} //
Phys.\ Rev.\ Lett. 2004. V.~92. P.~155003.

21.~{\it Robicheaux~F., Hanson~J.D.} //
Phys.\ Rev.\ Lett. 2002. V.~88. P.~055002.

%%%%%%%%%%%%%%%%%%%%%%%%%%%%%%%%%%%%%%%%%%%%%%%%%%%%%%%%%%%%

\end{document}